\documentclass[%
 aapm,
 mph,%
 amsmath,amssymb,
 reprint,%
]{revtex4-2}

\usepackage{graphicx}
\usepackage{svg}
\usepackage{float}
\usepackage{mathptmx}
\usepackage{dcolumn}
\usepackage{bm}
\usepackage{hyperref}
\usepackage{subfigure}
\usepackage[english]{babel}
\addto{\captionsenglish}{%
  
}

\begin{document}

\preprint{}

\title{Molecular Fluid Flow in MoS$_2$ Nanoporous Membranes and Hydrodynamics Interactions}

\author{Jo\~ao P. K. Abal}
\email{joao.abal@ufrgs.br}
\affiliation{Institute of Physics, Federal University of Rio Grande do Sul, 91501-970, Porto Alegre, Brazil}

\author{Marcia C. Barbosa}
\email{marcia.barbosa@ufrgs.br}
\affiliation{Institute of Physics, Federal University of Rio Grande do Sul, 91501-970, Porto Alegre, Brazi}

\date{\today}

\begin{abstract}
The shift in water security demands improvements in alternative solutions such as saltwater desalination. One of the most efficient technologies in this scope is the reverse osmosis systems, a technology based on a membrane separation process. MoS$_2$ nanoporous membranes are gained attention as a promise for the next-generation high selective and permeable membranes technology. Besides that, one aspect of nanoconfined fluid flow not yet investigated but studied from the fluid mechanics calculations is the impact of induced pressure fields in the water flux in neighboring microfilters, described as hydrodynamic interactions. For this purpose, we studied the water flow through adjacent MoS$_2$ nanopores by running Non-Equilibrium Molecular Dynamics simulations and obtained that in this scale the hydrodynamics interactions are not significant as expected.
\end{abstract}


\maketitle
\section{\label{sec:intro}introduction}

Water scarcity is one of the major challenges of our time. Changing climate patterns responsible for disturbing the hydrological cycle combined with growing water demand are shifting the water security towards high-risk levels~\cite{water}. In the face of the problem, seawater desalination technology has gained attention. Over the past decades, improvements in the sector have allowed a considerable reduction of power needed to desalinate seawater, due to advances in membrane technology and energy recovery equipment~\cite{VOUTCHKOV20182,Alvarez2018}. 

High-performance membranes, that can exhibit superior selectivity and high water flowrate are in the sight of the development of the next-generation desalination technology~\cite{Werber2016,Alvarez2018}. Meanwhile, computational models have been used to better understand the desalination process in the nanoscale. Molecular dynamics simulations are a powerful theoretical approach to study
the physics behind nanofluidic systems once it allows for probing the microscopic behavior of a collection of atoms while performing timescale feasible simulations~\cite{COHENTANUGI201559,Couette}. Through it, we can propose new membrane materials nanostructured designed to improve the desalination process.


One suited simulation branch to better understand the desalination process is mimic the reverse osmosis desalination system at the nanoscale. A pressure-driven transport can be created and the resulting water flow through nanopores and ions rejection can be studied~\cite{zhu-2014,li-acsnano2016,kohler-jcp2018,kou-pccp2016,cohen-graphene,C8CP02076K,doi:10.1021/la4018695,doi:10.1021/acs.nanolett.5b04089,aluru-mos2,Farimani-etal,D0CP00484G,doi:10.1063/1.4892638,doi:10.1021/acsnano.7b04299}. This technique enters in the scope of Non-Equilibrium Molecular Dynamics (NEMD). Also, its procedure has been used to get insights in design new membrane materials for desalination. Among the 2D membranes recently been investigated, the molybdenum disulfide (MoS$_2$) nanoporous membrane are a promising one~\cite{Farimani-etal,aluru-mos2,Environmental,li-nl2019,wang-nl2017,doi:10.1021/acsnano.7b05124,doi:10.1063/1.5104309,doi:10.1021/acs.jpcc.7b05153}.

\begin{figure}
\includegraphics[width=8cm]{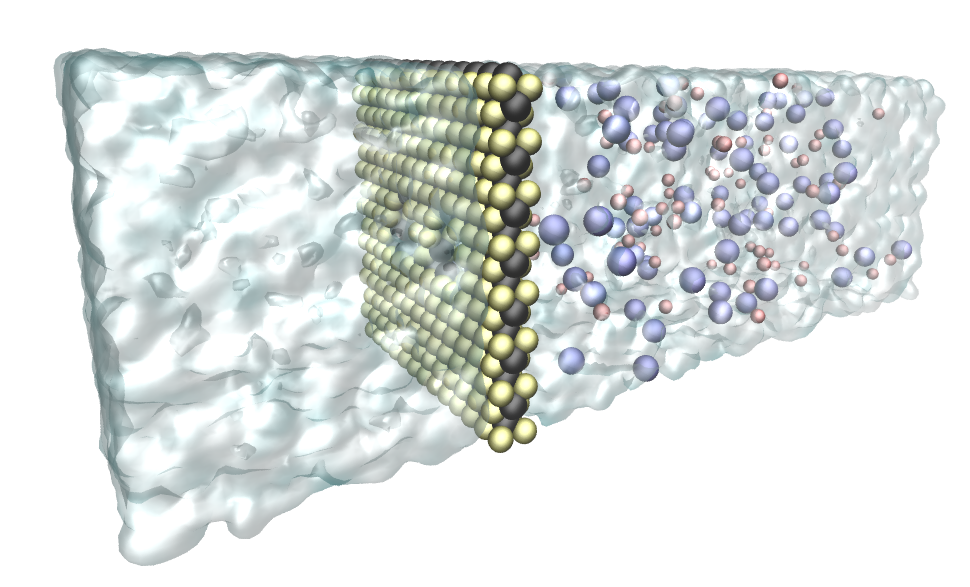}
\caption{\label{fig1}The illustration of a typical NEMD desalination system at the nanoscale. The saltwater (right side) is separated from the pure water (left side) by a MoS$_2$ nanoporous membrane. Pressure-driven transport can be simulated by imitating the reverse osmosis process. Image created using the VMD software~\cite{HUMP96}.}
\end{figure}

Usually, the water properties in confined systems differ from the bulk values. Also, the water transport mechanism confined in nanopores can be very different from the continuum hydrodynamics description~\cite{C3RA40661J}. The continuum hypothesis is one of the fundamentals assumptions of fluid mechanics, which is successful in describing the macroscopic behavior of fluid flow and states that fluid properties, such as pressure, density, and velocity, are well defined
at infinitesimally small points and vary continuously from one point to another~\cite{kannam_daivis_todd_2017}. However, in narrow nanopores (< 2 nm of diameter), the water flow is layered and a non-quadratic velocity profile emerges from it~\cite{C9CP04364K}. For such small molecular size pores, it is more useful to discuss the fluid transport using permeability and flow rate rather than viscosity and slip length, for example~\cite{kannam_daivis_todd_2017}.

Nevertheless, the fluid mechanics calculations in microfilters assume the existence of hydrodynamic interactions between adjacent pores. The interaction arises from the pressure field induced by the next pore which in turn makes the single pore water flow solution not sufficiently precise to expand its conclusions to the microfilters flow system~\cite{microfilters}. The influence of the pore number and its distance plays an important role in the overall water flux in the classical hydrodynamic picture. In addition, the simulations conducted so far in the scope of molecular dynamics desalination systems assume that the water flux results scale linearly with the nanopore number~\cite{C3EE43221A}, but assuming hydrodynamic interactions would lead to a deviation of this assumption if the nanopores are close enough. These open questions are elucidated in this work.

\begin{figure}[H]
\includegraphics[width=8cm]{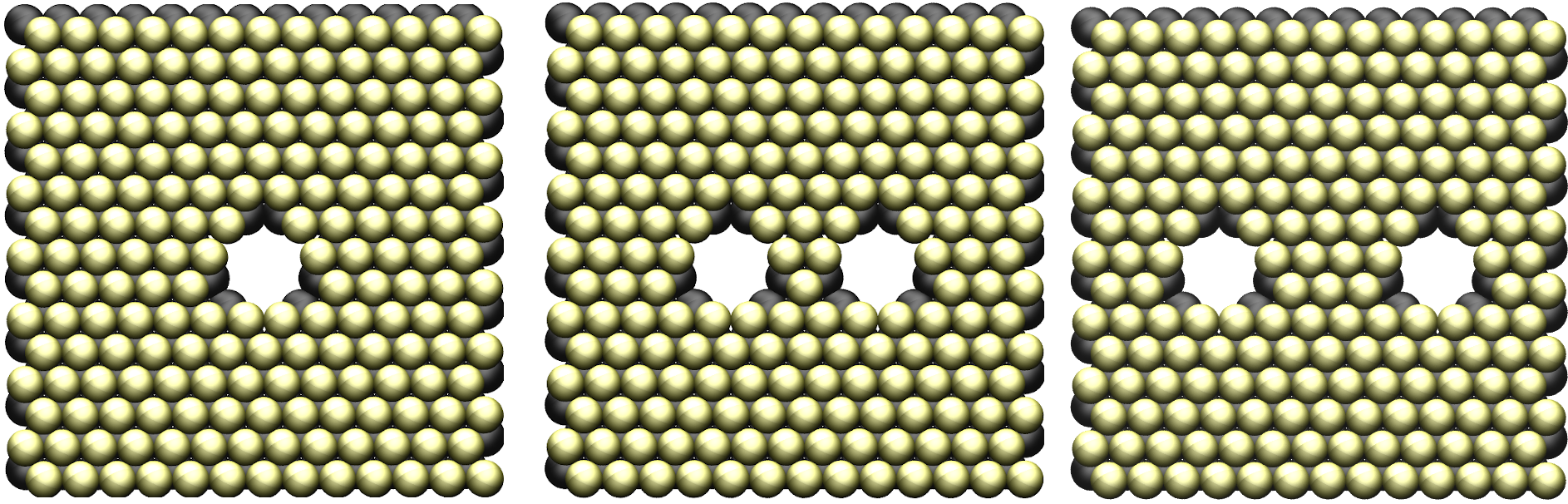}
\caption{\label{fig2}The MoS$_2$ nanoporous membranes studied in this work.}
\end{figure}

NEMD simulations were been conducted to shed light on the behavior of liquids in the nanoscale~\cite{PhysRevE.102.033110}. For the purpose of investigating if hydrodynamics interactions play a role in the nanoscale, the nanopore number and its proximity implications in water flux and salt rejection were evaluated in this work in the boundary of NEMD. To do so, we used three different MoS$_2$ membrane designs with different nanopore number (Figure~\ref{fig2}-left and others) and different nanopore distance (Figure~\ref{fig2}-center and right). These membranes were designed in order to maintain the nanopore chemistry and geometry the same in each case. So, the only difference in water flow would be due to hydrodynamic interactions.

\section{Computational Details}

The reverse osmosis desalination system can be designed in the scope of MD simulation as composed of two reservoirs (one of saltwater, that is the feed side, and another of pure water, that is the permeate side) separated by a membrane, as illustrated in Figure~\ref{fig1}. The two reservoirs can be confined by graphene barriers, for example, which in turn can serve as pistons to control the fluid pressure during the running desalination process.

\begin{table}[H]
\caption{\label{table1}The Lennard-Jones parameters and atoms charges employed in the simulations.}
\begin{ruledtabular}
\begin{tabular}{c c c c }
 & $\sigma_{LJ}$ [\AA] & $\varepsilon_{LJ}$ [kcal/mol] & Charge~(e) \\
\hline
    Na~\cite{nacl_marcia} & 2.52 & 0.0346 & 0.885 \\
    Cl~\cite{nacl_marcia} & 3.85 & 0.3824 & -0.885  \\
    O-Tip4p/$\varepsilon$~\cite{tip4p_marcia} & 3.165 & 0.1848 & -1.054 \\
    H-Tip4p/$\varepsilon$~\cite{tip4p_marcia} & 0.0 & 0.0 & 0.5270 \\
    Mo~\cite{KADANTSEV2012909} & 4.20 & 0.0135 & 0.6 \\
    S~\cite{KADANTSEV2012909} & 3.13 & 0.4612 & -0.3 \\
    C~\cite{Hummer2001} & 3.40 & 0.0860 & 0.0 \\
\end{tabular}
\end{ruledtabular}
\end{table}

As initial conditions in our simulations, the pure water side contains 1550 water molecules and the saltwater side contains 170 ions and 4930 water molecules, resulting in a solution of 1 mol/L of concentration. The MoS$_2$ membrane has a dimension of 4 x 4 nm and is held fixed in space. By doing that, we can work with high gradient pressures for statistical purposes. This is important in the sense of generating a sufficient number of events in a short time interval of 10 ns. The nanopores sizes were chosen to have 0.97 nm in diameter (defined as the distance center to center of atoms) in order to satisfy the minimum size in which the models employed does not show the ion blockage effect, as previous studies shown~\cite{D0CP00484G}.

The simulations were performed using the LAMMPS~\cite{lammps}. The particles interact with each other via Lennard-Jones (LJ) and Coulomb potential. The parameters used in this work are summarized in Table~\ref{table1}. The Tip4p/$\varepsilon$ and NaCl/$\epsilon$ model was used once they were parameterized to provide the correct value of bulk water dielectric constant~\cite{tip4p_marcia} and mixture dielectric constant~\cite{nacl_marcia}.

The simulations can be understood as follows, the first part has some equilibrations steps and the second is the non-equilibrium running process. In the first part, the two reservoirs were not in contact with each other, that is the nanopore is closed until the system is equilibrated properly. The simulations start with the total system energy been minimized during 0.5 ns in NVE ensemble. After that, a NPT ensemble was conducted during 1 ns at 300 K and 1 bar in each reservoir. Then, the simulation was further equilibrated for 2 ns at 300 K in NVT ensemble to achieve the water equilibrium density of 1 g/cm$^3$. Finally, the nanopore was opened by removing the desired atoms and the different pressures were applied in each reservoir for 10 ns. To achieve a steady-state flow, the pressure difference between the feed reservoir and the permeate one needs to overcome the osmotic pressure of the system, which in turn acts in the opposite direction. The feed pressures range from 100, 500, 1000, 2500, 5000 to 10000 bars.

\section{Results}

Usually, the water flux throughout the membrane is described by the quantity called membrane specific permeability~\cite{C3EE43221A}, which incorporates information about the nanopore density and the membrane resistance to water flow (the pressure needed to induce certain flow). 

\begin{figure}[H]
\includegraphics[width=8cm]{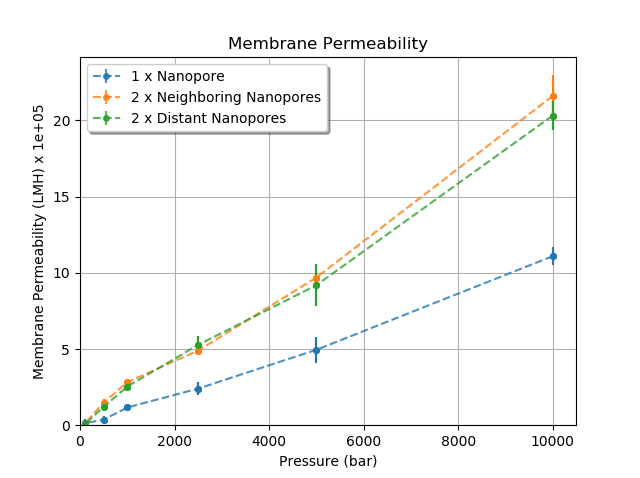}
\caption{\label{fig3}The membrane permeability as a function of pressure for each membrane design. The error bars are the deviation from the mean value.}
\end{figure}

The membrane specific permeability A$_m$ follows the expression: A$_m$ = $\phi$/(P - $\Pi$), in which $\phi$ is the water flux, $P$ is the applied pressure and $\Pi$ is the osmotic pressure, and has dimensions of $L/m^2/hr/bar$ or $LMH/bar$. The membrane permeability as a function of pressure for each membrane design is presented in Figure~\ref{fig3}. As expected, the membrane permeability is a linear function of the gradient pressure. The overall results for this specific membrane design are summarized in Table~\ref{table2}.

\begin{figure}[H]
\includegraphics[width=8cm]{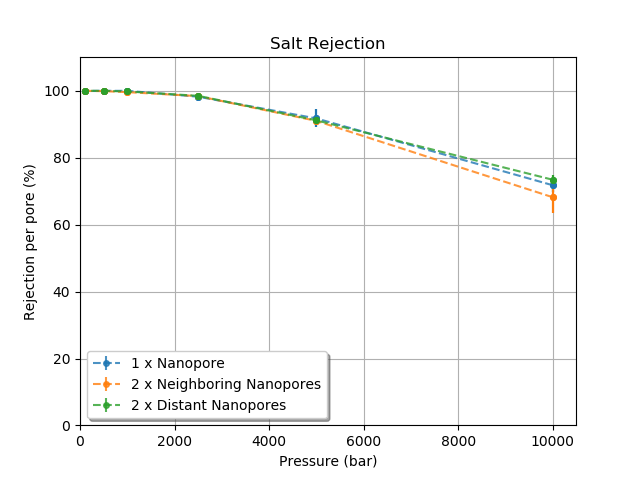}
\caption{\label{fig4}The salt rejection per nanopore as function of pressure. The error bars are the deviation from the mean value. For small pressure gradients (< 100 bars), near from the realistic process operation in reverse osmosis systems~\cite{doi:10.1063/1.4892638}, the salt rejection is $100\%$ for such MoS$_2$ nanopore size.}
\end{figure}

The salt rejection is the core of the desalination process. In reverse osmosis systems, usually, the salt rejection must be higher than 99\%~\cite{VOUTCHKOV20182}. The MoS$_2$ nanoporous membrane studied shows an excellent salt rejection capability, achieving 100\% of rejection per pore working in pressures below 1000 bar (Figure~\ref{fig4}). For a matter of comparison, the lowest pressure applied in this work was 100 bar, which in turn is higher than the real pressures used in reverse osmosis systems but it is justifiable in the scope of computational simulation for statistics purposes to generate a sufficient number of events in a time interval of 10 ns. As we can see in Figure~\ref{fig4}, the salt rejection per pore is not affected by the nanopore proximity or number.

\begin{table}[H]
\caption{\label{table2}The membrane specific permeabilities (A$_m$) obtained as function of nanopore density and distance. The numbers inside the parentheses are the membrane specific permeabilities standard deviations evaluated in this work.}
\begin{ruledtabular}
\begin{tabular}{c c c}
\textbf{Nanopore Density [10$^{12}$~cm$^{-2}$]} & A$_m$ [LMH/bar] & Distance [nm] \\
\hline
    1 x Nanopore - 6.25  & 101.7 (25.2) &  \\
    2 x Neighboring Nanopores - 12.5 & 242.9 (55.8) & 1.275\\
    2 x Distant Nanopores - 12.5 & 223.6 (38.1) & 1.913\\
\end{tabular}
\end{ruledtabular}
\end{table}

As we know, in a general way the water flux ($Q$) is a function of the water density inside the pore channel ($\rho$), the water velocity through it ($V$), and the pore area ($A$), that is $Q = \rho \cdot V \cdot A$. The area $A$ of the pores is a geometric parameter that, in turn, is maintained constant in our simulations. The density $\rho$ and the axial velocity $V$ are the remaining control parameters and they are related to the pore chemistry~\cite{aluru-mos2,Farimani-etal,COHENTANUGI201559}. The pore chemistry depends on the particle interactions and their distribution around the pore. We know from previous studies that the charge distribution affects the overall water flux~\cite{C4RA02856B,doi:10.1021/jp400578u,doi:10.1063/1.5003695}. In our simulations, we choose such an arrangement of atoms, as illustrated in Figure~\ref{fig2}, to maintain constant the proportion between hydrophobic and hydrophilic sites. As a consequence of this choice, the charge distribution is the same in each case and the pores are charged neutral. In summary, the nanopore chemistry and geometry are the same in all simulations. By doing that, we expect that any change in the water flux as a function of nanopore number or distance would be due to hydrodynamics interactions between the pores, which in turn would be reflected in the water flux or water density around the pores~\cite{microfilters}.

\begin{figure}
\includegraphics[width=8cm]{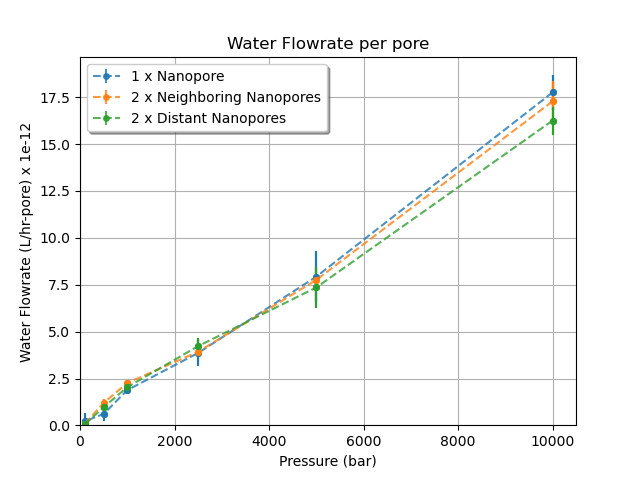}
\caption{\label{fig5}Water flowrate per pore as a function of applied pressure.}
\end{figure}

However, as we can see from Figure~\ref{fig5}, the water flowrate per pore does not appear to have any dependence on nanopore number or distance. To clarify this question, we investigate the water density inside and around the pore to see if the influence would be hidden or compensated by other factors.

First, we classify some regions of analysis, as illustrated in Figure~\ref{fig6}. The \textit{Region 1} is defined by two water layers near the membrane in the permeate side, which corresponds approximately to a slice of 5 \AA~in z-direction. The \textit{Region 2} is defined as a slice of 2 \AA~ between the nanopore region and the first two water layers representing the \textit{Region 1}. The \textit{Region 3} is defined by the nanopore region.

\begin{figure*}
\includegraphics[width=2\columnwidth]{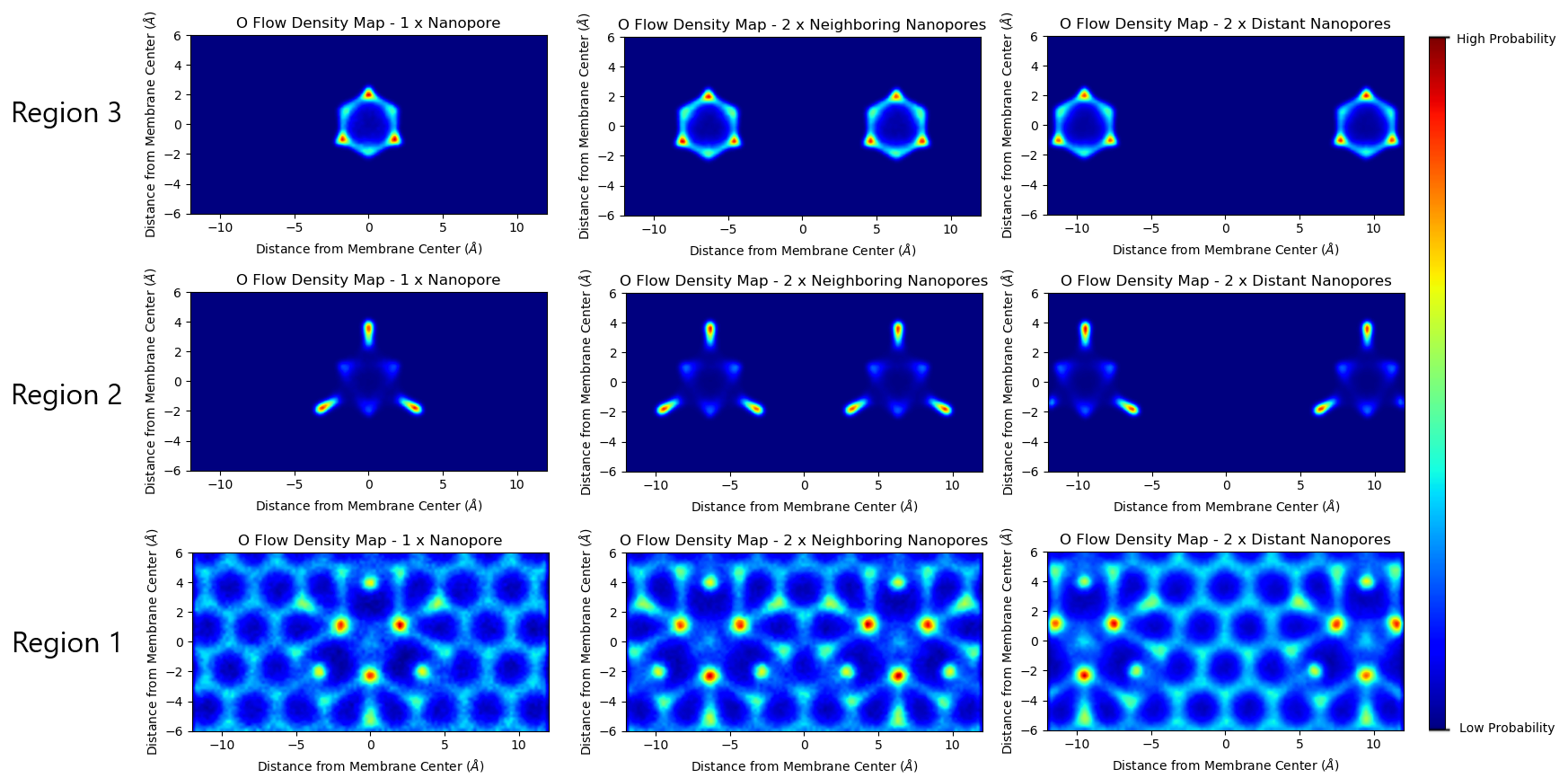}
\caption{\label{fig7}Oxygen density map averaged over all simulations shown for each region of analysis.}
\end{figure*}

\begin{figure}[H]
\includegraphics[width=8cm]{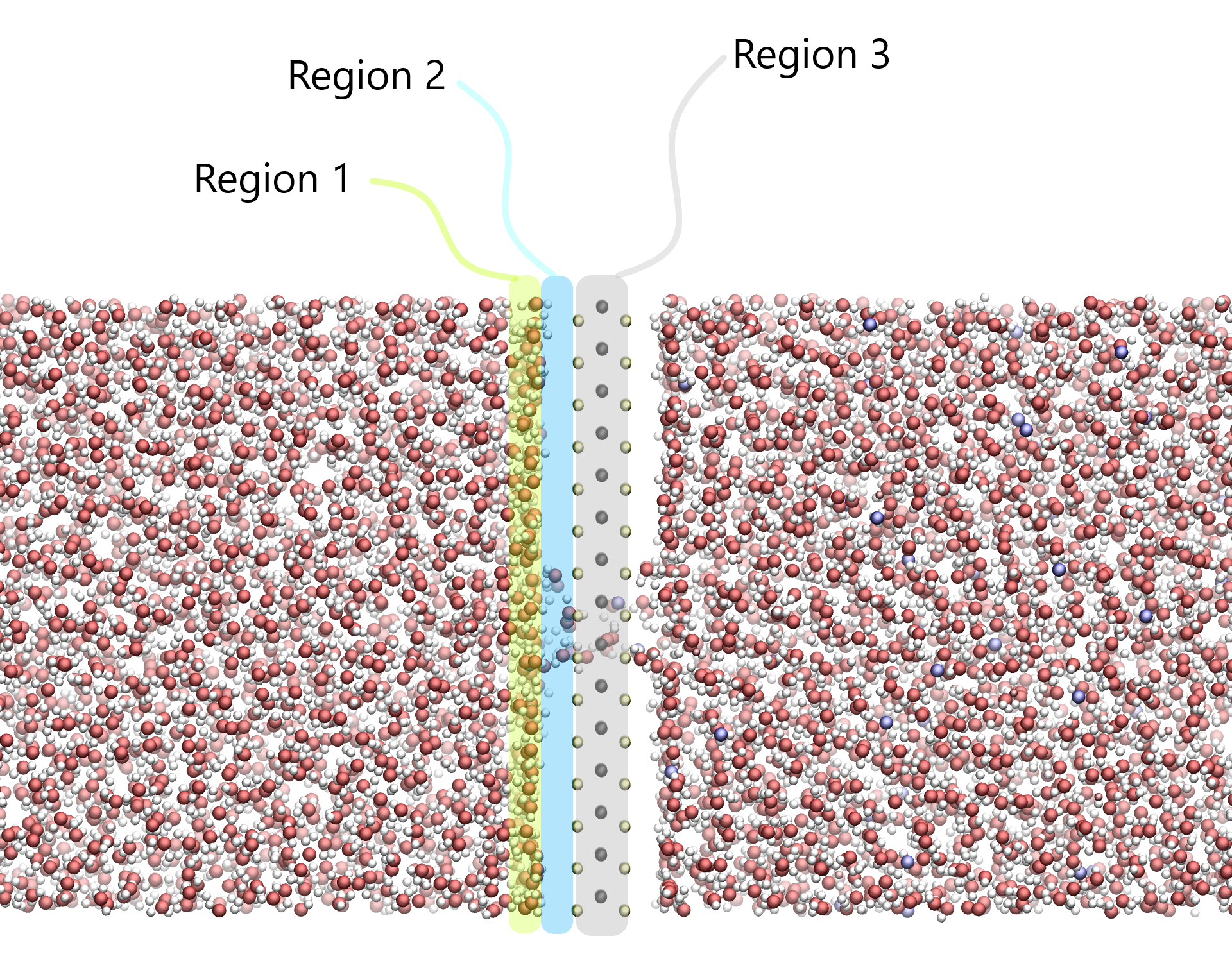}
\caption{\label{fig6}Definition of the oxygen density map regions of analysis.}
\end{figure}

The oxygen density map from Figure~\ref{fig7}-\textit{Region 3} in the single nanopore case (left column) shows how the water moves through it. As it can be seen, the water molecules transport occurs in some specific regions around the center of the nanopore and not in the real center. Layered water structure, described by density oscillations in the radial direction, arises and it is a signature of the implications of nanoconfinement. Also interesting is the fact that the red-shifted areas are the highest ones in terms of occupancy and it corresponds to the sites near the Mo atoms. From Figure~\ref{fig7}-\textit{Region 2} it can be seen that water molecules enter in the nanopore attracted mostly by the Mo sites, as previous studies confirmed~\cite{aluru-mos2}. In addition, the first two water layers in Figure~\ref{fig7}-\textit{Region 1} show that in this slice the oxygen of water molecules prefer to stay between S sites, which is in fact the region in which the Mo-water electrostatic interaction is less screened by the S atoms. 

The oxygen density map from Figure~\ref{fig7}-\textit{Region 1} in the single nanopore case (left column) shows that the first two water layers are modified locally by the presence of the nanopore. However, its extension is not larger than the nanopore size of 0.97 nm of diameter, which suggests that the nanopore presence does not have a large effect in the water structure near the membrane, just local implications near the nanopore region.

\begin{figure}[H]
\includegraphics[width=8cm]{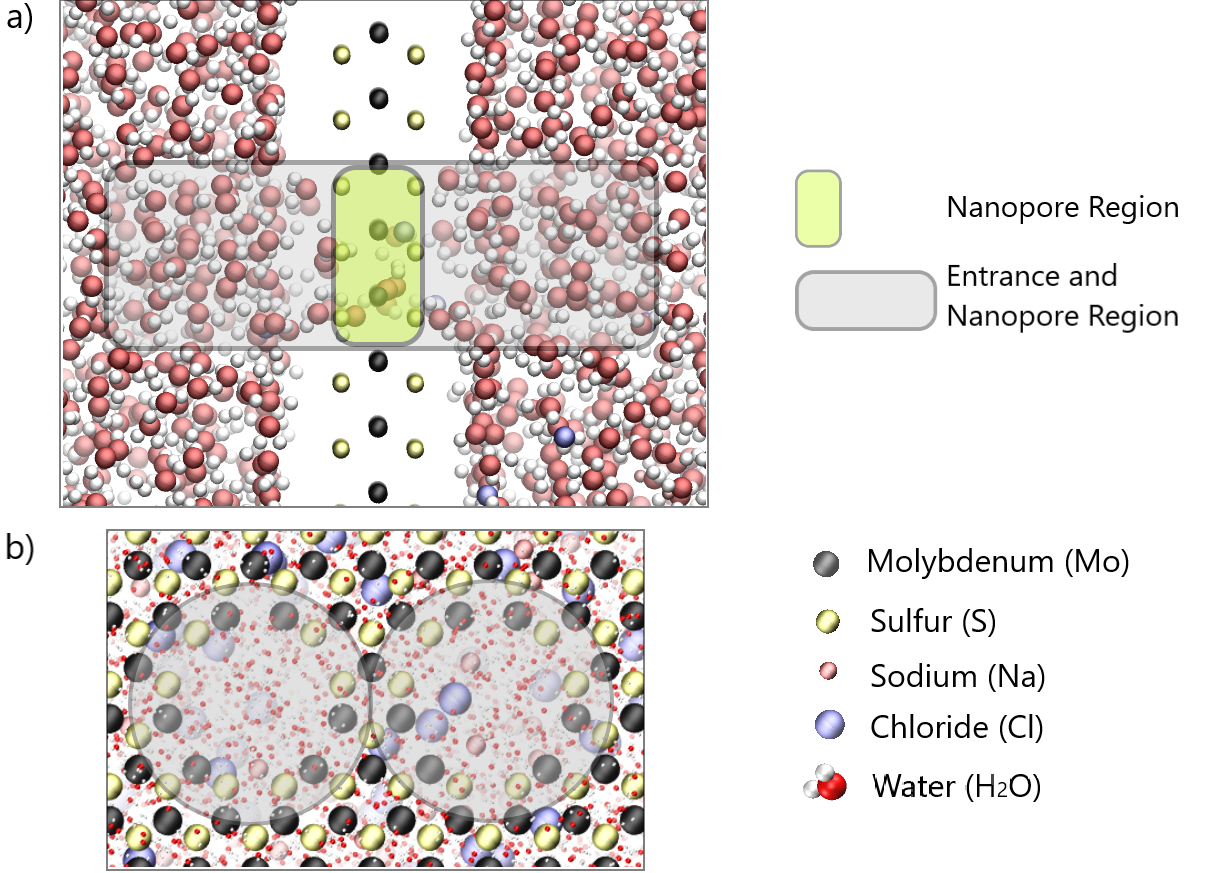}
\caption{\label{fig8}a) The definitions of nanopore region and entrance region for the water density analysis. b) The front view of the cylindrical regions of analysis.}
\end{figure}

Comparing with the two neighboring nanopores case, we can't see any deviation in the density map due to the presence of the second nanopore. This conclusion extends to the third case, the distant nanopores case. To quantify if any implications in the water density exist due to the proximity of nanopores, we obtained the water density as a function of the radial distance from each nanopore center, as defined in Figure~\ref{fig8}. The radial water density was calculated binning the region inside the nanopore in circular shapes, counting the water molecules there, and dividing by its cylindrical volume.

\begin{figure}[H]
\includegraphics[width=8cm]{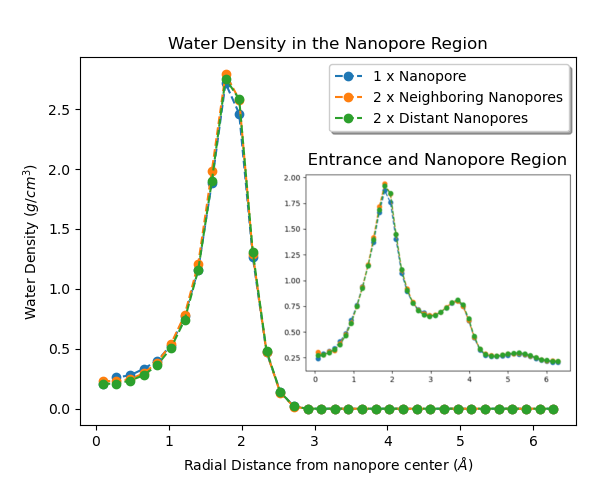}
\caption{\label{fig9}The water density as a function of radial distance from each nanopore center in the nanopore region. In the detail: The water density as a function of radial distance from each nanopore center in the entrance and nanopore region.}
\end{figure}

As we can see from Figure~\ref{fig9} there is no difference in the water density inside the nanopore due to the presence of a second one. In addition, the water density is related to the Potential of Mean Force (PMF) through the integration of the following expression: $F = - RTln[\rho]$~\cite{Farimani-etal,doi:10.1021/acs.jpcc.7b06480}. If any induced pressure field extends from one nanopore to the other one, it is not sufficient to produce any change in the water density inside the nanopore and as a consequence in the water PMF. Not only inside the nanopore but also when we investigate the entrance region, as illustrated in Figure~\ref{fig9} in the detail, we do not observe any significant effect.

The hydrodynamic effects were not visible in this scale may be due to the polar nature (atomic charges) of the MoS$_2$ membrane, which induces its structure in the firsts water layers~\cite{doi:10.1021/acs.jpcc.7b05153}.

\section{Conclusions}

Differently from the fluid mechanics calculations in microfilters~\cite{microfilters}, the fluid flow through neighboring nanopores in MoS$_2$ membranes does not show in our NEMD simulations any significant hydrodynamic interactions between adjacent pores. The water flow strongly depends on the intermolecular force of the membrane, which is governed by the layered structure of the liquid in the nanopore region, and as a consequence, the collective effect of hydrodynamic interaction between pores is suppressed. Nevertheless, we shed light on the assumption that the water flux would scale linearly with the nanopore density regardless of its distance. Of course, here the MoS$_2$ atoms were held fixed in space, and more careful simulations are needed to understand the relation between nanopores distance and material strain. As previous studies confirmed, the MoS$_2$ nanoporous membranes are promising candidates for the next-generation membrane material, allowing water to be filtered at high permeate rates while maintaining high salt rejection rates.

\bibliographystyle{unsrt}
\bibliography{aapmsamp}

\end{document}